\documentstyle[12pt,epsfig]{article}
\newcommand{\Frac}[2]%
{{\textstyle \frac{\mbox{\footnotesize $#1$}\rule[-0.9mm]{0mm}{1mm}}%
{\mbox{\footnotesize $#2$}\rule{0mm}{3.1mm}}}}
\renewcommand{\thefootnote}{\fnsymbol{footnote}}

\begin{document}
\begin{titlepage}
\vspace*{-12 mm}
\noindent
\begin{flushright}
\begin{tabular}{l@{}}
\end{tabular}
\end{flushright}
\vskip 12 mm
\begin{center}
{\large \bf 
Spin constraints on Regge predictions and perturbative evolution 
in high energy collisions 
}
\\[14 mm]
{\bf Steven D. Bass}
\\[10mm]   
{\em
Institute for Theoretical Physics, \\
Universit\"at Innsbruck,
Technikerstrasse 25, Innsbruck, A 6020 Austria
\\[5mm]
}
\end{center}
\vskip 10 mm
\begin{abstract}
\noindent
Two key issues in the application of perturbative QCD and Regge 
predictions to high energy processes 
are whether the hard and soft pomerons should be considered as 
two separate distinct exchanges
and whether the Regge intercepts are $Q^2$ independent or not.
Models involving a distinct hard pomeron exchange predict much 
larger values for the LHC total cross-section.
Here we argue that there is a polarized analogue of this 
issue in the isovector part of the spin structure function $g_1$
and that the spin data appear to favour a distinct hard exchange.
\end{abstract}

\vspace{9.0cm}

\\


\end{titlepage}
\renewcommand{\labelenumi}{(\alph{enumi})}
\renewcommand{\labelenumii}{(\roman{enumii})}
\renewcommand{\thefootnote}{\arabic{footnote}}

\newpage
\baselineskip=6truemm

Regge and small $x$ physics has been vigorously studied in the
context of  HERA and is important for predicting the LHC total
cross-section.
There are various models and approaches which depend on whether
the hard pomeron observed at
HERA
should be treated as a distinct exchange
\cite{twopomerons}.
Predictions for the LHC total cross-section range
from about 90 mb up to about 150 mb
\cite{lhcsigma,pvl05}
with the larger values associated with a distinct hard-pomeron
\cite{pvl05}.
In this paper we address the questions:
Can we use present information from polarized processes
\cite{bassrmp}
to help constrain models of Regge and small $x$ dynamics and
is there evidence in polarized data for distinct hard exchanges ?

Models of small $x$ physics generally fall into two clases.
First consider unpolarized scattering.
The first approach \cite{convention}
involves a soft-pomeron
and perturbative QCD evolution
(DGLAP, $\alpha_s^n \ln^m \frac{1}{x}$ ... resummation)
which
drives the increase in the effective intercept $\alpha$,
$F_2 \sim x^{-\alpha}$,
from the soft-pomeron value 0.08 to the value $\sim 0.4$
extracted from HERA data \cite{hera}.
In a second approach Cudell et al. \cite{Cudell:1999} have argued
that the Regge intercepts should be independent of $Q^2$
and that the HERA data is described by a distinct
hard-pomeron exchange in addition to the
soft-pomeron.
The hard pomeron should also appear in low $Q^2$ photoproduction data
and in proton-proton collisions.
There are two conflicting measurements of the total cross-section at
the Tevatron \cite{fnal}.
The larger CDF measurement favours a separate hard pomeron contribution.

Is there a place in polarized data where similar physics issues occur ?
There are interesting clues in the data.
Our knowledge of the $g_1$ spin structure function at deep inelastic $Q^2$
mostly comes
from
SLAC small $x$ data between 0.01 and 0.1 for the isovector part
of $g_1$
\cite{slac}
and
COMPASS small $x$ data between 0.004 and 0.1 for the isoscalar part
\cite{compass}.
The isovector part of $g_1$ rises as $\sim x^{-0.5}$
and
is much bigger than the isoscalar part of $g_1$,
which is close to zero in the measured kinematics in sharp
contrast to the unpolarized structure function \cite{soffer,bassrmp,bassmb}.
In this paper we focus on the isovector part of $g_1$.

The rise in $g_1^{p-n} = g_1^p - g_1^n$
is a challenge for Regge predictions and
perturbative QCD.
The Regge prediction for $g_1^{p-n}$ at small $x$ is
\begin{equation}
g_1^{p-n} \sim \sum_i f_i \ \biggl( \frac{1}{x} \biggr)^{\alpha_i}
.
\end{equation}
Here the $\alpha_i$ denote the Regge intercepts
for
isovector $a_1$ Regge exchange
and the $a_1$-pomeron cuts \cite{Heimann:1973}.
The coefficients $f_i$ are to be determined from experiment.
If one makes the usual assumption that the $a_1$ Regge trajectories
are straight lines parallel to the $(\rho, \omega)$
trajectories then one finds
$\alpha_{a_1} \simeq -0.4$
for the leading trajectory,
within the phenomenological range $-0.5 \leq \alpha_{a_1} \leq 0$
discussed in Ref.~14. 
Taking the masses of the $a_1(1260)$ and $a_3(2070)$
states
together with the $a_1(1640)$ and $a_3(2310)$ states
from the
Particle Data Group \cite{PDG:2004}
yields two parallel $a_1$ trajectories with slope
$\sim 0.75$GeV$^{-2}$
and a leading trajectory with slightly lower intercept:
$\alpha_{a_1} \simeq -0.18$.
For this value of $\alpha_{a_1}$
the effective
intercepts corresponding to the $a_1$ soft-pomeron cut and
the $a_1$ hard-pomeron cut are $\simeq -0.1$ and $\simeq +0.22$
respectively
{\it if} one takes the
soft and hard pomerons as two distinct exchanges
\footnote{
I thank P.V. Landshoff for valuable discussions on this issue.
}
Values of $\alpha_{a_1}$ close to zero could be achieved with curved
Regge trajectories; the recent model of Brisudova et al.~\cite{goldman}
predicts $\alpha_{a_1} = -0.03 \pm 0.07$.
For this value the intercepts of the $a_1$ soft-pomeron cut and
the $a_1$ hard-pomeron cut become $\sim +0.05$ and $\sim +0.37$.
The $a_1$ and $a_1$ soft-pomeron cut alone are unable to account
for the data.
\footnote{
It should be noted that, in the measured $x$ range,
the effective isovector
exponent 0.5 could be softened through multiplication by a
$(1-x)^n$ factor --
for example associated with perturbative QCD counting
rules at large $x$ ($x$ close to one).
For example, the exponent $x^{-0.5}$ could be modified
to about
$x^{-0.25}$ through multiplication by a factor $(1-x)^6$.
However, this is not sufficient to reconcile the measured
rising structure function
with
the naive Regge prediction involving soft $a_1$ exchange.
}

The observed rise in $g_1^{p-n}$ at deep inelastic values of
$Q^2$ is required to reproduce the area under the fundamental Bjorken
sum rule \cite{Bjorken:1966}
\begin{eqnarray}
& &
\int_0^1 dx g_1^{(p-n)} (x,Q^2)
\nonumber \\
& &
\ \ \ \ \
=
\frac{g_A^{(3)}}{6}
\biggl[ 1
- \frac{\alpha_s(Q^2)}{\pi}
- 3.583 \left(\frac{\alpha_s(Q^2)}{\pi} \right)^2
- 20.215 \left(\frac{\alpha_s(Q^2)}{\pi} \right)^3 + ... \biggr]
\end{eqnarray}
Here $g_A^{(3)} = 1.2695 \pm 0.0029$ \cite{PDG:2004}
is the scale-invariant
isovector axial-charge measured in neutron beta-decays.
The sum-rule has been confirmed in polarized deep inelastic scattering
experiments at the level of 10\%
\cite{bjsr}.
About 50\% of the Bjorken sum-rule comes from $x$ values less than about
0.12 and 10\% comes from $x$ values less than 0.01
\cite{bassrmp,soffer,slac}.
The $g_1^{p-n}$ data is consistent with quark model and
perturbative QCD
predictions in the valence region $x > 0.2$ \cite{epja}.
The size of $g_A^{(3)}$ forces us to accept a large contribution
from small $x$
and the observed rise
in $g_1^{p - n}$ is required to fulfil this non-perturbative constraint.

Does this rise follow from $a_1$ exchange plus perturbative QCD evolution
or is there a distinct hard exchange ?
-- that is, a polarized analogue of the one {\it or} two pomerons question!
The difference between the effective intercept describing
$g_1^{p-n}$ at deep inelastic values of $Q^2$ and the prediction based
on soft $a_1$ exchange is a factor of up to 2-3 bigger
than the difference in
the effective intercept needed to describe $F_2$
in the unpolarized HERA data and the soft-pomeron prediction.

In the conventional approach the $a_1$ term (or $a_1$ soft-pomeron cut)
should describe the high-energy part of $g_1$ close to photoproduction
and provide the input for perturbative QCD evolution at deep inelastic
values of $Q^2$ above the transition region.
One then applies perturbative QCD
(DGLAP or DGLAP
 plus double logarithm $\ln^2 \frac{1}{x}$ ... resummation)
and out should come
the rising structure function seen in the data
\cite{blumlein,ziaja,bartels,ermolaev}.
For $g_1^{p-n}$
with DGLAP evolution
this approach has the challenging feature that
the input and output (at soft and hard scales)
are governed by non-perturbative constraints with perturbative QCD
evolution in the middle
unless the $a_1$ Regge input has information about $g_A^{(3)}$
built into it.
(Furthermore, perturbative $\alpha_s^{l+1} \ln^{2l}x$ resummation
 calculations predict a sharp rise
 \cite{ziaja,bartels,ermolaev}
 $\sim x^{-\gamma}$ with $\gamma \sim 0.9-1$
 in the absolute value of
 the isosinglet spin structure function $g_1^d$
 which is not observed in the present data.
 The measured structure function is consistent with zero for $x$
 between the lowest value 0.004 and 0.05 \cite{compass},
 and the integral
 $\int_{x_{\rm min}}^1 dx g_1^d$ is observed
 to converge within the errors for $x_{\rm min} \sim 0.05$ \cite{hermes}.)
The alternative scenario is a separate hard-exchange contribution
(perhaps an $a_1$ hard-pomeron cut) in addition to the soft $a_1$.

Some guidance may come from looking at the QCD evolution equations
in moment space.
We first consider the perturbative DGLAP approach since this is
presently the prime tool used
to analyse polarized deep inelastic data.
To test deep inelastic
sum-rules it is necessary to have all data points at the same value
of $Q^2$. In the experiments the different data points are measured
at different values of $Q^2$, viz. $x_{\rm expt.}(Q^2)$.
Next-to-leading order (NLO) QCD-motivated fits taking into account
the scaling violations associated with perturbative QCD are
used to evolve all the data points to the same $Q^2$
using DGLAP evolution
\cite{compass,evolution}.
The results of these fits are then extrapolated to $x \sim 0$ to test
spin sum-rules.

Let $\Delta q_3 (x,t) = (\Delta u - \Delta d)(x,t)$
denote the isovector spin-dependent parton distribution with
$t = \ln \frac{Q^2}{\Lambda^2}$;
$\int_0^1 dx \Delta q_3 (x,t) = g_A^{(3)}$.
The DGLAP equation for $\Delta q_3 (x,t)$ is
\begin{equation}
\frac{d}{dt} \Delta q_3 (x,t)
=
\int_x^1 \frac{dy}{y} P \biggl( \frac{x}{y} \biggr) \Delta q_3 (y,t)
\end{equation}
where
\begin{equation}
P (z) =
\frac{\alpha_s(t)}{2 \pi}
C_2 (R)
\biggl[ \frac{1 + z^2}{(1-z)_+} + \frac{3}{2} \delta(z-1) \biggr]
\end{equation}
is the leading-order spin-dependent splitting function;
$C_2 (R) = \frac{4}{3}$
and
$P (z)$ goes to a constant as $z \rightarrow 0$.
The area under
$\Delta q_3 (x,t)$ is conserved because of the Bjorken sum-rule.
DGLAP evolution \cite{altp} acts to shift
the weight of the distribution and $g_1^{p-n}$ to smaller $x$
with increasing $Q^2$,
meaning that a convergent input will be unstable to DGLAP evolution
at a given value of small $x$
\cite{ABFR}
and prompting the question at what values of $Q^2$ and small $x$
should spin-dependent Regge predictions work in this approach,
if any ?

The evolution equation (3) becomes ``singular'' in the $x \rightarrow 0$
limit if $\Delta q_3$ behaves as a constant for $x \rightarrow 0$:
\begin{equation}
{\rm convolution} \sim \int_x^1 \frac{dy}{y} \biggl\{ ... \biggr\}
.
\end{equation}
This compares with the singlet channel in unpolarized scattering
where the splitting matrix has a
$1/z$ singularity as $z$ goes to zero for evolution into gluons.
If the unpolarized gluon distribution were to have a leading $1/y$
pole then the contribution
\begin{equation}
P \biggl( \frac{x}{y} \biggr) g(y) \sim \frac{y}{x} . \frac{1}{y}
\end{equation}
would yield the same structure in the evolution equation.

Take the Mellin transform $\int_0^1 dz z^{N-1} P(z)$:
\begin{equation}
P (N, \alpha_s(Q^2))
=
\int_0^1 dz z^{N-1} P (z)
\nonumber \\
=
\frac{\alpha_s}{2 \pi}
C_2 (R)
\biggl[
- \frac{1}{2} + \frac{1}{N(N+1)} - 2 \sum_{j=2}^{N} \frac{1}{j} \biggr]
.
\end{equation}
The zeroth moment of the DGLAP splitting function has a pole at $N=0$
at leading order (LO)
plus higher-order poles at NLO \cite{vogelsang}.
{\it If} we require that the scattering amplitude is analytic
in $Q^2$ \cite{eden},
then Regge singularities are independent of $Q^2$ and
new singularities should not suddenly appear as $Q^2$ increases.
This result has practical consequences for
singularities in the Mellin transform of the DGLAP splitting function:
the $N=0$ poles
become an artifact of the perturbative expansion.
That is, they should vanish in a full (non-perturbative)
resummation
\cite{Cudell:1999}
otherwise one will generate an unphysical fixed pole $\alpha =0$
contribution in
the isovector part of $g_1$ as soon as one reaches large enough $Q^2$
to apply DGLAP evolution.
To see this, consider the Mellin transform of the spin dependent
parton distribution
\begin{equation}
u (N, Q^2) = \int_0^1 dx x^{N-1} \Delta q_3 (x, Q^2/\Lambda^2)
\end{equation}
and its DGLAP equation
\begin{equation}
\frac{\partial}{\partial t} u (N, Q^2)
= P  (N, \alpha_s(Q^2)) u(N, Q^2)
.
\end{equation}
If the twist-two term
$u(N,Q^2)$ has no pole at $N=0$ at $Q^2$ values close to photoproduction,
then the solution to the DGLAP equation
\begin{equation}
u(N,Q^2) = \exp \biggl[ \
C \ \log \frac{\log (Q^2 / \Lambda^2)}{\log (Q^2_0 / \Lambda^2) }
\ P(N) \ \biggr]
u(N,Q_0^2)
\end{equation}
automatically generates an essential singularity in $u(N,Q^2)$
at $N=0$
as soon as $Q^2$ becomes large enough for the application of
perturbative QCD;
$C = 6 / (33 - 2f)$ where $f$ is the number of active flavours.
This is not allowed if we assume that $u(N,Q^2)$
is analytic at $N=0$ for some finite range of $Q^2$ \cite{Cudell:1999,eden}.
It cannot suddenly acquire a fixed singularity at $N=0$ when it is
analytically continued in $Q^2$.
To help understand $P(N)$,
Donnachie and Landshoff \cite{pvlpol}
consider the example of the analogous expansion of
the function
$
\psi (N, \alpha_s) = \sqrt{N^2 + \alpha_s} - N
= \frac{\alpha_s}{2 N} - \frac{\alpha_s^2}{8 N^3} + ...
$.
Although each term in the expansion is singular at $N=0$,
the function
$\psi$ is not: the expansion is valid only for $|N| > \alpha_s$.
(Related issues in unpolarized scattering are discussed in
 Ref.~32
 where a new small $x$ splitting function
 has been proposed with no $1/N$ pole in the Mellin transform.)

The first moment of $P(z)$ vanishes, $P(1, \alpha_s)=0$, corresponding
to the conserved axial-charge $g_A^{(3)}$.
The positive odd moments of the DGLAP splitting function correspond
to the anomalous dimensions of axial-tensor operators in
the light-cone operator product expansion for deep inelastic scattering.
There are no operators corresponding to the poles at $N=0$ or $N=-1$.

Going beyond DGLAP evolution, Bl\"umlein and Vogt \cite{blumlein}
have considered
the resummation of $\alpha_s^{l+1} \ln^{2l} x$
terms in the evolution kernels of non-singlet contributions
to $g_1$.
An all orders resummation
of these terms in perturbation theory leads only to corrections of
1\% for $g_1^{p-n}$ relative to NLO calculations in the kinematics
accessible
to present experiments.
The most singular contributions in the perturbation expansion
behave like a power series in $N (\alpha_s/N^2)^k$
when we take the Mellin transform and work with the moments.
One assumes that $\alpha_s < 3 \pi N^2 / 8$
-- see Eq.(13) of Ref.~21 
with the number of colours $N_c=3$.
Like for the DGLAP procedure discussed above,
each term in the perturbative expansion is singular for $N=0$ in moment
space.
One again encounters the issue of whether the isovector
$g_1^{p-n}$ structure function is analytic at $N=0$ for finite $Q^2$.

Motivated by this discussion we consider a hard exchange ``input''
to perturbative evolution.
For a fixed power behaviour
$
u(x,t) \sim f(Q^2) x^{- \epsilon}
$
the Mellin transform is
$
u (N, Q^2) \sim \frac{ f(Q^2)}{N - \epsilon }
.
$
Substituting this into the DGLAP equation and equating the coefficient of
the pole gives an equation for the coefficent of the Regge exponent:
\begin{equation}
\frac{\partial}{\partial t} f (Q^2)
=
P (N= \epsilon, \alpha_s(Q^2)) \ f(Q^2)
.
\end{equation}
If there is a hard exchange with fixed intercept
away from the pole at
$\epsilon =0$,
e.g. $\epsilon = 0.5$
or perhaps $\sim 0.2$
for the $a_1$ hard-pomeron cut
(plus $(1-x)^n$ counting rules factors still at work in the measured
 $x$ range),
then
a combined
Regge-DGLAP approach should be a good approximation
--
just as a distinct hard pomeron would resolve challenging issues
in the interpretation of the unpolarized structure function.
(A rising ``input'' $g_1^{p-n} \sim x^{-0.2}$ was used
 in Ref.~21 
at the input scale
 $Q_0^2 = 4$ GeV$^2$,
 safely away from the pole at $\epsilon=0$.)
Further, if the intercept is $Q^2$ independent the issue of
reconciling the Regge input
to perturbative QCD
evolution and the Bjorken sum rule constraint would be resolved.
The hard exchange could be looked for in low $Q^2$ data -- see below.

In the isosinglet sector it is harder to draw firm conclusions.
$g_1^{p+n}$ is small and consistent with zero in the measured
small $x$ kinematics \cite{compass}.
The Regge prediction involves a
contribution
$\sim \{ 2\ln \frac{1}{x} -1 \}$
from two non-perturbative gluon exchange \cite{Bass:1994,Close:1994}
plus
contributions from the $f_1$ trajectory and $f_1$-pomeron cuts.
It is unknown whether the gluon exchange contribution Reggeizes
or whether it is a fixed pole.
Brodsky et al. \cite{brodbs}
have argued that colour coherence
forces $\Delta g (x) / g(x) \propto x$ when $x \rightarrow 0$.
In this scenario we might also expect a polarized version of
the hard pomeron with intercept $\sim +0.4$ which would
correspond to a rising term (in absolute value)
like $x^{-0.4}$ as $x \rightarrow 0$.
Perhaps the coefficients of these terms are separately suppressed
or perhaps they cancel in the measured kinematics ?

To summarize, {\it if} we assume analyticity in $Q^2$
then the isovector spin structure function $g_1^{p-n}$
favours a hard exchange contribution at small $x$ with
a $Q^2$ independent Regge intercept.
This exchange should also contribute to and could be looked for
in high-energy polarized photoproduction and
in the transition region between $Q^2=0$ and deep inelastic
values of $Q^2$ ($Q^2 < 1$GeV$^2$).
High-energy polarized photoproduction and the transition region
could be investigated using a polarized electron-proton collider
\cite{bassadr}
or perhaps through measurement of low $Q^2$ asymmetries at COMPASS
using a proton target.
A hard exchange contribution might also show up in the spin-dependent
part of the proton-proton total cross-section.
In polarized proton-proton collisions one would be looking
for a leading behaviour
$\Delta \sigma \sim s^{-0.5}$ to $\sim s^{-0.8}$
instead of
the simple
$a_1$ prediction $\sim s^{-1.4}$
and
non-perturbative gluon-exchange contribution
$\sim (\ln s/\mu^2) / s$
with $\mu \sim 0.5 - 1$~GeV
a typical hadronic scale \cite{Bass:1994,Close:1994}.
Will these processes exhibit evidence of a hard
exchange with Regge intercept
$\alpha \sim +0.5$ or just the exchanges predicted
by soft Regge theory ?
These spin measurements,
together with the total cross-section at the LHC,
would help constrain our understanding of hard and soft exchanges in
high energy collisions.

\section*{Acknowledgments}

I thank W. Guryn and P.V. Landshoff for helpful conversations.
This research is supported by the Austrian Science Fund (grant P17778-N08).

\end{document}